\documentclass[aip,amsmath,amssymb,twocolumn,]{revtex4}                

\usepackage{textcase}
\usepackage{graphicx}
\usepackage{dcolumn}
\usepackage{bm}
\usepackage{amsmath,amsthm,amssymb,mathrsfs}

\begin{document}

\title{Reduction of back switching by large damping ferromagnetic material}

\author{
  Tomohiro Taniguchi${}^{1}$, Yohei Shiokawa${}^{2}$, and Tomoyuki Sasaki${}^{2}$
} 

\affiliation{ 
 ${}^{1}$National Institute of Advanced Industrial Science and Technology (AIST), Research Center for Emerging Computing Technologies, Tsukuba, Ibaraki 305-8568, Japan, \\
 ${}^{2}$TDK Corporation, Advanced Products Development Center, Ichikawa, Chiba 272-8558, Japan
}


\begin{abstract}
Recent studies on magnetization dynamics induced by spin-orbit torque have revealed a weak dependence of the critical current for magnetization switching on the damping constant of a ferromagnetic free layer. 
This study, however, reveals that the damping constant nevertheless plays a key role in magnetization switching induced by spin-orbit torque. 
An undesirable switching, returning to an initial state, named as back switching, occurs in a ferromagnet with an easy axis parallel to the current direction. 
Numerical and theoretical analyses reveal that back switching is strongly suppressed when the damping constant of the ferromagnet is large. 
\end{abstract}

\maketitle


Low-damping ferromagnetic materials has been investigated for spintronics applications \cite{oogane06,bilzer06,mizukami08,iihama12,konoto13,tsunegi14}. 
These materials are interesting because the critical current for exciting magnetization dynamics by spin-transfer torque \cite{slonczewski96,berger96} in two-terminal devices is typically proportional to the damping constant \cite{sun00,grollier03}, 
and therefore, low-damping ferromagnetic materials help to reduce power consumption. 
The value of the damping constant in typical ferromagnets, such as Fe, Co, Ni, and their alloys, is on the order of $10^{-3}-10^{-2}$, 
even after the effect of spin pumping \cite{tserkovnyak02PRL,mizukami02JMMM} is taken into account \cite{konoto13,tsunegi14}. 
Studies on three-terminal devices manipulated by spin-orbit torque \cite{liu12,pai12,yu14,cubukcu14,you15,torrejon15,fukami16,fukami16NM,lau16,brink16,oh16,shiokawa19}, 
however, have revealed that the dependence of the critical current on the damping constant is weak 
when the easy axis of the ferromagnet is perpendicular to the film plane \cite{lee13,taniguchi19} or parallel to the current direction \cite{taniguchi20}; 
these systems are named type Z and type X, respectively, in Ref. \cite{fukami16}. 
Then, a question arises as to what the role of the damping constant is in three-terminal devices. 


The purpose of this study is to investigate the relation between the magnetization state manipulated by the spin-orbit torque and the magnetic damping constant by solving the Landau-Lifshitz-Gilbert (LLG) equation. 
We focus on the type-X system because the switching mechanism in the system is not fully understood yet, unlike the other systems \cite{sun00,grollier03,lee13,taniguchi19}. 
An undesirable return of the magnetization, called back switching in this paper, occurs in a ferromagnet, 
where, although the spin-orbit torque brings the magnetization close to the switched state, it returns to the initial state after turning off the current. 
As a result, the phase diagram of the magnetization state as a function of the current and external magnetic field alternately shows switched and non-switched states. 
It is revealed that the back switching appears as a result of the magnetization precession around the perpendicular axis after the current is turned off. 
The back-switching region is strongly suppressed by using large-damping ferromagnetic materials. 



\begin{figure}
\centerline{\includegraphics[width=1.0\columnwidth]{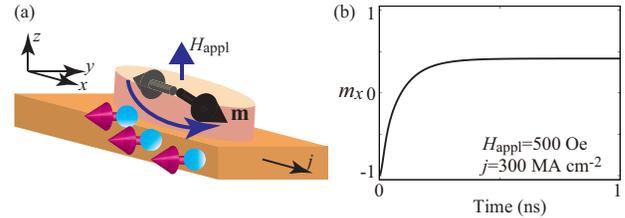}}
\caption{
         (a) Schematic illustration of the system. 
         The spin Hall effect in the bottom nonmagnet generates spin current with a polarization along the $y$ direction 
         and excites the magnetization ($\mathbf{m}$) dynamics in the top ferromagnet. 
         The easy axis of the ferromagnet is parallel to the $x$ axis. 
         An external magnetic field $H_{\rm appl}$ is applied in the $z$ direction. 
         (b) Example of the time evolution of $m_{x}$ for $j=300$ MA cm${}^{-2}$ and $H_{\rm appl}=500$ Oe. 
         \vspace{-3ex}}
\label{fig:fig1}
\end{figure}




The system studied in the work is an in-plane magnetized ferromagnet placed on a nonmagnetic metal 
and schematically shown in Fig. \ref{fig:fig1}(a). 
Electric current flowing in the nonmagnet in the $x$ direction generates pure spin current that is injected into the ferromagnet with the spin polarization in the $y$ direction. 
The spin current excites the spin-transfer torque acting on the magnetization in the ferromagnet and induces magnetization dynamics. 
The $z$ axis is perpendicular to the film plane. 
We denote the unit vectors pointing in the magnetization direction in the ferromagnet and along the $k$-axis ($k=x,y,z$) as $\mathbf{m}$ and $\mathbf{e}_{k}$, respectively. 
The magnetization dynamics are described by the LLG equation 
\begin{equation}
  \frac{d\mathbf{m}}{dt}
  =
  -\gamma
  \mathbf{m}
  \times
  \mathbf{H}
  -
  \gamma
  H_{\rm s}
  \mathbf{m}
  \times
  \left(
    \mathbf{e}_{y}
    \times
    \mathbf{m}
  \right)
  +
  \alpha
  \mathbf{m}
  \times
  \frac{d\mathbf{m}}{dt},
  \label{eq:LLG}
\end{equation}
where $\gamma$ and $\alpha$ are the gyromagnetic ratio and the Gilbert damping constant, respectively. 
The magnetic field $\mathbf{H}$ consists of 
the in-plane magnetic anisotropy field $H_{\rm K}$ in the $x$ direction, the demagnetization field $-4\pi M$ in the $z$ direction, 
and an external field $H_{\rm appl}$ in the $z$ direction: 
\begin{equation}
  \mathbf{H}
  =
  H_{\rm K}
  m_{x}
  \mathbf{e}_{x}
  +
  \left(
    H_{\rm appl}
    -
    4\pi M 
    m_{z}
  \right)
  \mathbf{e}_{z}. 
\end{equation}
The strength of the spin-transfer torque is given by 
\begin{equation}
  H_{\rm s}
  =
  \frac{\hbar \vartheta j}{2eMd}, 
\end{equation}
where $\vartheta$ is the spin Hall angle in the nonmagnet, 
whereas $M$ and $d$ are the saturation magnetization and thickness of the ferromagnet. 
The electric current density is denoted as $j$. 
The values of the parameters used in this study were taken from typical experiments \cite{liu12,pai12,yu14,cubukcu14,you15,torrejon15,fukami16,fukami16NM,lau16,brink16,oh16} 
as $M=1500$ emu c.c.${}^{-1}$, $H_{\rm K}=200$ Oe, $\gamma=1.764\times 10^{7}$ rad Oe${}^{-1}$ s${}^{-1}$, $\vartheta=0.4$, and $d=1.0$ nm. 


The magnetic field $\mathbf{H}$ is related to the magnetic energy density $E$ via 
$E=-M \int d\mathbf{m}\cdot\mathbf{H}$, 
\begin{equation}
  E
  =
  -MH_{\rm appl}
  m_{z}
  -
  \frac{MH_{\rm K}}{2}
  m_{x}^{2}
  +
  2\pi M^{2}
  m_{z}^{2}.
  \label{eq:energy}
\end{equation}
The energy density $E$ has two minima at $\mathbf{m}_{0\pm}=\pm\sqrt{1-m_{0z}^{2}}\mathbf{e}_{x}+m_{0z}\mathbf{e}_{z}$, where $m_{0z}=H_{\rm appl}/(H_{\rm K}+4\pi M)$. 
Throughout this paper, the initial state is set to be $\mathbf{m}_{0-}$, which points in the negative $x$ direction. 
Accordingly, we call the other stable state, $\mathbf{m}_{0+}$ pointing in the positive $x$ direction, the switched state. 
Thus, we are interested in experiments where the initial state is reset to $\mathbf{m}_{0-}$ during each trial of magnetization switching. 
By convention, we will focus on switching by a positive current and field region. 



\begin{figure}
\centerline{\includegraphics[width=1.0\columnwidth]{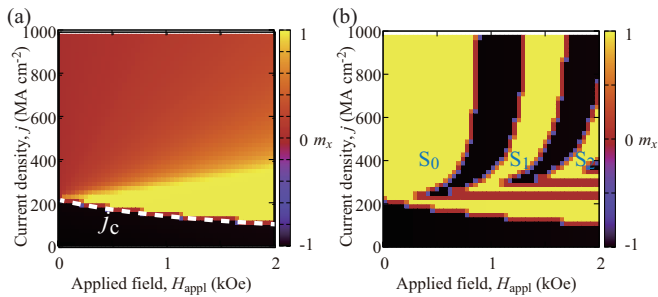}}
\caption{
         Phase diagrams of $m_{x}$ (a) at a fixed point in the presence of current 
         and (b) in a relaxed state after turning off the current. 
         The initial state is $\mathbf{m}_{0-}$. 
         The white dotted line in (a) represents the theoretical formula of the critical current density derived in Ref. \cite{taniguchi20}. 
         Switching regions are distinguished by the labels S${}_{n}$ ($n=0,1,2$) in (b). 
         The damping constant $\alpha$ is $0.005$. 
         \vspace{-3ex}}
\label{fig:fig2}
\end{figure}




Figure \ref{fig:fig1}(b) shows an example of the time evolution of $m_{x}$ in the presence of current. 
The damping constant is $\alpha=0.005$. 
The results indicate that the magnetization saturates to a fixed point. 
Figure \ref{fig:fig2}(a) is a phase diagram summarizing the fixed point of $m_{x}$ in the presence of current, 
where the vertical and horizontal axes represent the current density $j$ and the external magnetic field $H_{\rm appl}$. 
The magnetization stays near the initial state [$m_{x}\simeq -1$ shown in black in Fig. \ref{fig:fig2}(a)] in a relatively small current region, 
where its boundary is well explained by the critical-current formula $j_{\rm c}$ \cite{taniguchi20} 
shown by the white dotted line in Fig. \ref{fig:fig2}(a). 
On the other hand, the magnetization state above the critical current satisfies $m_{x}>0$. 
Therefore, one might suppose that the magnetization relaxes to the switched state, $m_{x}\simeq +1$, after turning off the current. 


However, the phase diagram after turning off the current shown in Fig. \ref{fig:fig2}(b) reveals a complicated dependence of the relaxed state on the current and external magnetic field. 
There are stripes distinguishing the switched ($\mathbf{m}_{0+}$, yellow) and non-switched ($\mathbf{m}_{0-}$, black) states. 
This result indicates that the magnetization returns to the initial state under certain conditions. 
We name this phenomenon back switching and investigate its relation to the damping constant in the following. 


Let us briefly comment on another phenomenon, called backhopping \cite{sun09,oh09,skowronski13,abert18,safranski19}, to avoid any confusion. 
Backhopping is a phenomenon in which, after magnetization switching in a free layer is achieved, an undesirable return to the original resistance-state occurs at a high-bias voltage. 
There are two differences between back switching and backhopping. 
First, backhopping in two-terminal devices originates from magnetization switching in the reference layer. 
On the other hand, back switching relates to the magnetization dynamics in the free layer. 
Second, backhopping has often been investigated by sweeping the current. 
On the other hand, we reset the initial state of the magnetization at each trial 
because we are interested in designing the operation conditions of memory devices. 
Back switching originates from the fact that the fixed point of the magnetization in the presence of current is not parallel to the easy axis of the magnetization, as discussed below, 
and thus, it will occur in not only type-X devices, but also type-Z devices \cite{lee13,taniguchi19}. 


Note that the horizontal stripe in the relatively low current region of Fig. \ref{fig:fig2}(b), slightly above $j_{\rm c}$, was analyzed in Ref. \cite{taniguchi20}, 
and therefore, it will be excluded from the following discussion. 
We label the other switched regions shown by the vertical stripes as the S${}_{n}$-region ($n=0,1,2,\cdots$); see Fig. \ref{fig:fig2}(b). 
The role of the integer $n$ will be clarified below. 




\begin{figure}
\centerline{\includegraphics[width=1.0\columnwidth]{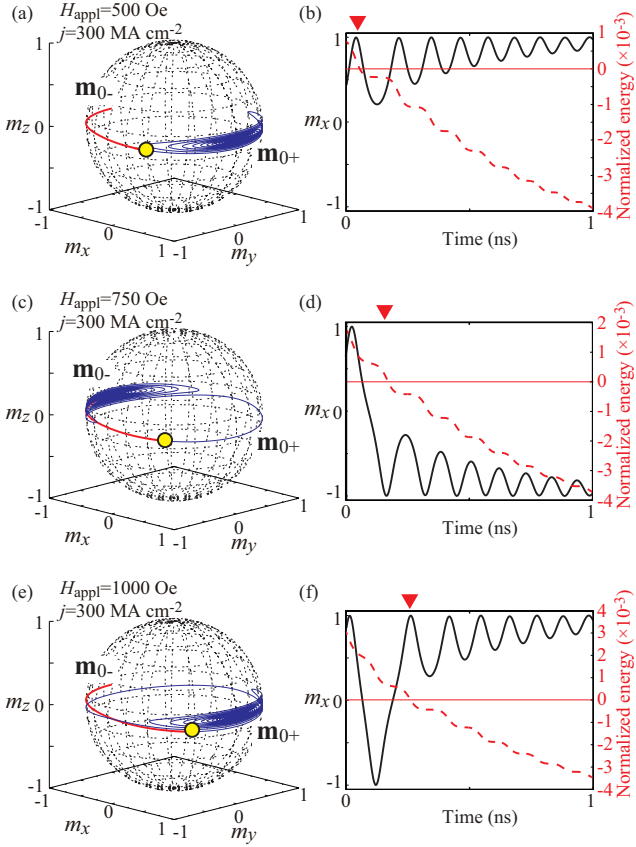}}
\caption{
         (a) Dynamic trajectory in the S${}_{0}$-region; $H_{\rm appl}=500$ Oe, and $j=300$ MA cm${}^{-2}$. 
             The red line represents the trajectory from the initial state to the fixed point (yellow circle) in the presence of the current, 
             whereas the blue line corresponds to the relaxation dynamics after turning off the current. 
             The symbols $\mathbf{m}_{0\pm}$ indicate the locations of the initial ($\mathbf{m}_{0-}$) and switched ($\mathbf{m}_{0+}$) states. 
         (b) Time evolutions of $m_{x}$ (black solid line) and normalized energy density $\varepsilon$ (dashed red line) after turning off the current. 
             The red triangle indicates the time at which $\varepsilon$ becomes zero. 
         (c) Dynamic trajectory and (d) time evolution of $m_{x}$ and $\varepsilon$ in the region sandwiched by the S${}_{0}$- and S${}_{1}$-regions; $H_{\rm appl}=750$ Oe, and $j=300$ MA cm${}^{-2}$. 
         (e) Dynamic trajectory and (f) time evolution of $m_{x}$ and $\varepsilon$ in the S${}_{1}$-region; $H_{\rm appl}=1000$ Oe, and $j=300$ MA cm${}^{-2}$. 
         \vspace{-3ex}}
\label{fig:fig3}
\end{figure}



Figure \ref{fig:fig3}(a) shows the dynamic trajectory of the magnetization obtained by numerically solving Eq. (\ref{eq:LLG}), 
where $j=300$ MA cm${}^{-2}$ and $H_{\rm appl}=500$ Oe correspond to the S${}_{0}$ region. 
The red line represents the trajectory from the initial state to the fixed point in the presence of the current, 
whereas the blue line shows the relaxation dynamics after turning off the current. 
The fixed point satisfying $d\mathbf{m}/dt=\bm{0}$ in the presence of the current is indicated by a yellow circle. 
Note that the fixed point is far away from the energetically stable states, $\mathbf{m}_{0\pm}$. 
This is because the spin-transfer torque drives the magnetization in the $y$ direction. 
The black solid line in Fig. \ref{fig:fig3}(b) shows the time evolution of $m_{x}$ after turning off the current. 
Starting from the fixed point located in the region of $m_{x}>0$, the magnetization relaxes to the switched state $\mathbf{m}_{0+}$ after showing several precessions around it. 
Figures \ref{fig:fig3}(c) and \ref{fig:fig3}(d), on the other hand, show an example of back switching, 
where the parameter $H_{\rm appl}=750$ Oe corresponds to the region sandwiched by the S${}_{0}$ and S${}_{1}$-regions. 
In this case, the magnetization after turning off the current shows a precession around the $z$ axis. 
As a result, even though the fixed point in the presence of the current is in the region of $m_{x}>0$, the magnetization returns to the initial state ($\mathbf{m}_{0-}$). 
Let us then move to the next switching region, the S${}_{1}$-region. 
The dynamic trajectory, as well as $m_{x}$, shown in Figs. \ref{fig:fig3}(e) and \ref{fig:fig3}(f), indicates that, 
starting from the fixed point in the region of $m_{x}>0$, the magnetization returns once to the region of $m_{x}<0$ before relaxing to the switched state. 
Now the meaning of the integer $n$ ($n=0,1,2,\cdots$) we used to distinguish the switched regions becomes clear: 
it represents how many times the magnetization returns to the region of $m_{x}<0$ before relaxing to the switched state. 


These results imply that the precession around the $z$ axis after turning off the current is the origin of back switching. 
Such a precession is induced by the precession torque, $-\gamma \mathbf{m}\times \mathbf{H}$. 
In particular, the precession around the $z$ axis occurs when the energy density at the fixed point of the magnetization in the presence of the current 
is larger than the saddle-point energy given by 
\begin{equation}
  E_{\rm d}
  =
  -\frac{MH_{\rm appl}^{2}}{8\pi M}.
  \label{eq:energy_saddle}
\end{equation}
In fact, a fixed point of the LLG equation is given by 
\begin{align}
  m_{x}
  =
  \frac{H_{\rm appl}H_{\rm s}}{H_{\rm s}^{2}-H_{\rm K}4\pi M},
&&
  m_{z}
  =
  -\frac{H_{\rm appl}H_{\rm K}}{H_{\rm s}^{2}-H_{\rm K}4\pi M}.
  \label{eq:fixed_point}
\end{align}
Substituting Eq. (\ref{eq:fixed_point}) into Eq. (\ref{eq:energy}), the energy density at this fixed point is 
\begin{equation}
  E_{j}
  =
  \frac{MH_{\rm appl}^{2}H_{\rm K}}{2(H_{\rm s}^{2}-H_{\rm K}4\pi M)}.
  \label{eq:energy_fixed_point}
\end{equation}
Note that $H_{\rm s}^{2}-H_{\rm K}4\pi M>0$ because the back switching appears in the current region above $j_{\rm c}$ for $H_{\rm appl}=0$, 
which is given by $j_{\rm c}=[2eMd/(\hbar\vartheta)]\sqrt{H_{\rm K}4\pi M}$ \cite{taniguchi20}. 
Therefore, Eq. (\ref{eq:energy_fixed_point}) is always positive, whereas the saddle-point energy density given by Eq. (\ref{eq:energy_saddle}) is negative. 
Accordingly, the fixed point in the presence of the current is in the unstable region corresponding to an energy larger than the saddle-point energy. 
As a result, after turning off the current, the magnetization starts to precess around the $z$ axis, as mentioned above. 
Simultaneously, the magnetization loses energy due to damping torque, 
where the dissipated energy density is given by 
\begin{equation}
  \Delta E(t) 
  =
  \frac{\alpha \gamma M}{1+\alpha^{2}}
  \int_{0}^{t}
  dt^{\prime} 
  \left[
    \mathbf{H}^{2}
    -
    \left(
      \mathbf{m}
      \cdot
      \mathbf{H}
    \right)^{2}
  \right].
  \label{eq:delta_E}
\end{equation}
Here, $t$ is the time after the current is turned off. 
When the condition, 
\begin{equation}
  E_{j}
  -
  \Delta E(t) 
  =
  E_{\rm d},
  \label{eq:relaxation_condition}
\end{equation}
is satisfied, the magnetization relaxes to the nearest stable state. 
The point here is that the magnetization alternately comes close to two stable states, $\mathbf{m}_{0\pm}$, due to the precession around the $z$ axis 
before Eq. (\ref{eq:relaxation_condition}) is satisfied. 
As a result, both $\mathbf{m}_{0+}$ and $\mathbf{m}_{0-}$ can be the final state, which leads to back switching. 
The conclusions of the discussion are confirmed by the dashed red lines in Figs. \ref{fig:fig3}(b), \ref{fig:fig3}(d), and \ref{fig:fig3}(f), 
where are the time evolutions of the energy density. 
Here, we introduce the normalized energy density, 
\begin{equation}
  \varepsilon(t)
  =
  \frac{E(t)-E_{\rm d}}{4\pi M^{2}}.
  \label{eq:energy_normalized}
\end{equation}
The energy density $E$ is a function of time with the initial condition of $E(t=0)=E_{j}$ 
because $E$, given by Eq. (\ref{eq:energy}), depends on the magnetization direction, 
and the magnetization changes direction in accordance with the LLG equation. 
The condition given by Eq. (\ref{eq:relaxation_condition}) is $\varepsilon(t)=0$ in terms of the normalized energy density. 
The times at which $\varepsilon$ becomes zero are indicated by the red triangles in the figures. 
It is clear from Figs. \ref{fig:fig3}(b), \ref{fig:fig3}(d), and \ref{fig:fig3}(f) that the final state of the magnetization is determined by whether $m_{x}$ is positive or negative when $\varepsilon$ becomes zero. 



\begin{figure}
\centerline{\includegraphics[width=1.0\columnwidth]{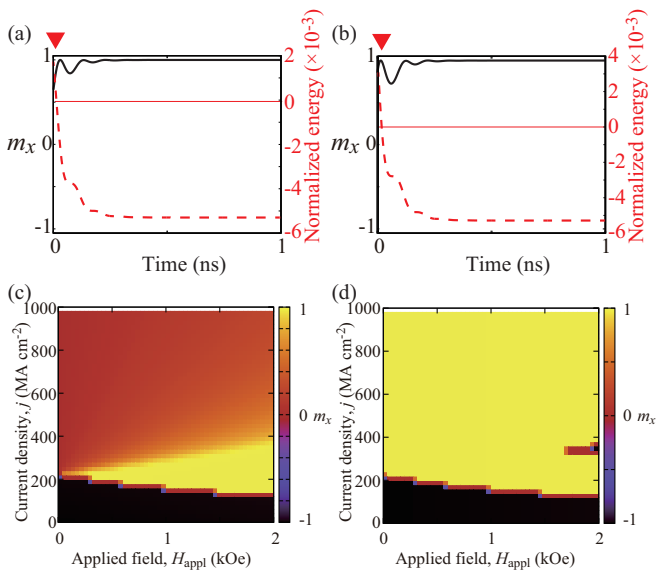}}
\caption{
         Time evolutions of $m_{x}$ (black solid lines) and normalized energy density $\varepsilon$ (red dashed lines) after turning off the current, 
         where $j=300$ MA cm${}^{-2}$ and (a) $H_{\rm appl}=750$ Oe and (b) $1000$ Oe. 
         The damping constant $\alpha$ is $0.050$. 
         The red triangles indicate the time at which $\varepsilon$ becomes zero. 
         Phase diagrams of $m_{x}$ 
         (c) at a fixed point in the presence of current and (d) in a relaxed state after turning off the current. 
         \vspace{-3ex}}
\label{fig:fig4}
\end{figure}



It is expected that using a large-damping ferromagnetic material will result in a reduction of the back-switching region. 
Remember that back switching occurs due to the precession around the $z$ axis. 
When the damping constant is large, the energy dissipation given by Eq. (\ref{eq:delta_E}) becomes large within a short time $t$, 
and the condition given by Eq. (\ref{eq:relaxation_condition}) is immediately satisfied before the magnetization returns to the region of $m_{x}<0$ by the precession around the $z$ axis. 
Accordingly, back switching does not occur. 
On the other hand, when the damping constant is small, it takes a long time to dissipate the energy to satisfy Eq. (\ref{eq:relaxation_condition}), 
during which time the magnetization shows the precession around the $z$ axis. 
As a result, back switching appears. 


To verify this picture, we evaluated the phase diagram of the magnetization state for a relatively large damping constant, $\alpha=0.050$, 
which is ten times large than that used above. 
Figures \ref{fig:fig4}(a) and \ref{fig:fig4}(b) show the time evolutions of $m_{x}$ and $\varepsilon$ after turning off the current, where $H_{\rm appl}$ is (a) $750$ Oe and (b) $1000$ Oe. 
Contrary to the dynamics shown in Figs. \ref{fig:fig3}(d) and \ref{fig:fig3}(f), 
the magnetization in Figs. \ref{fig:fig4}(a) and \ref{fig:fig4}(b) does not return to the region of $m_{x}<0$ because the energy dissipates quickly due to the large damping torque. 
As a result, the magnetization immediately relaxes to the switched state. 
Figures \ref{fig:fig4}(c) and \ref{fig:fig4}(d) summarize the magnetization state in the presence of current and after turning off the current, respectively. 
Comparing Fig. \ref{fig:fig4}(c) with Fig. \ref{fig:fig2}(a), it is clear that the phase diagram of the magnetization state in the presence of the current is approximately independent of the damping constant. 
On the other hand, the back-switching region is strongly suppressed, as can be seen by the comparing Fig. \ref{fig:fig2}(b) and \ref{fig:fig4}(d). 
This is because the large damping torque immediately dissipates the energy from the ferromagnet and leads to a fast relaxation to the switched state, 
due to which the S${}_{0}$-region dominates in the phase diagram. 


The existence of the back switching gives an upper limit of a write-current margin for memory applications, 
and therefore, it restricts the device design and/or manipulation conditions. 
On the other hand, back switching might be applicable to other devices such as a random number generator. 
The analysis in this study provides fruitful insights for the development of spintronics applications utilizing spin-orbit torque. 


In conclusion, the phase diagram of the magnetization state in a type-X spin-orbit torque device was calculated as a function of the electric current density and the external magnetic field. 
The magnetization state after turning off the current has stripe structures alternately showing switched and non-switched states. 
Such a non-switched state, named back switching here, occurs as a result of the magnetization precession around the perpendicular axis after the current is turned off. 
The back-switching region is reduced by using large-damping ferromagnetic materials, whereas the critical current inducing the magnetization switching is approximately independent of the damping constant. 


The authors thank to Masamitsu Hayashi, Shinji Isogami, Toru Oikawa, Takehiko Yorozu, and Seiji Mitani for valuable discussion. 
This work was supported by funding from TDK Corporation. 




\end{document}